\documentclass[conference,10pt]{IEEEtran}
\usepackage{amsfonts}
\usepackage{mathrsfs}
\usepackage{amssymb}
\usepackage{wrapfig}
\usepackage{amsmath}
\usepackage[top=0.75in, bottom=1in, left=0.625in, right=0.625in, textwidth=7.25in,textheight=9.25in]{geometry}
\usepackage{graphicx}
\usepackage{setspace}
\usepackage{url}
\usepackage{color}

\IEEEoverridecommandlockouts

\makeatletter
\def\ps@headings{%
\def\@oddhead{\mbox{}\scriptsize\rightmark \hfil \thepage}%
\def\@evenhead{\scriptsize\thepage \hfil \leftmark\mbox{}}%
\def\@oddfoot{}%
\def\@evenfoot{}}
\makeatother 

\begin{document}
\title{\huge\textbf{A Fast-CSMA Algorithm for Deadline-Constrained Scheduling over Wireless Fading Channels}}
\author{Bin Li and Atilla Eryilmaz\\Department
of Electrical and Computer Engineering\\The Ohio State University,
Columbus, Ohio 43210 USA.\\Email: \{lib, eryilmaz\}@ece.osu.edu}
\date{}


\maketitle

\newtheorem{theorem}{Theorem}
\newtheorem{lemma}{Lemma}
\newtheorem{claim}{Claim}
\newtheorem{definition}{Definition}
\newtheorem{assumption}{Assumption}
\newtheorem{remarks}{Remarks}

\newcommand{\CapReg}{\Lambda}
\newcommand{\cS}{\mathcal{S}}
\newcommand{\vS}{\mathbf{S}}
\newcommand{\vQ}{\mathbf{Q}}
\newcommand{\bE}{\mathbb{E}}

\def\QEDclosed{\mbox{\rule[0pt]{1.3ex}{1.3ex}}}
\def\QEDopen{{\setlength{\fboxsep}{0pt}\setlength{\fboxrule}{0.2pt}\fbox{\rule[0pt]{0pt}{1.3ex}\rule[0pt]{1.3ex}{0pt}}}}
\def\QED{\QEDopen}
\def\maximize{\operatornamewithlimits{Maximize}}
\def\limitsup{\operatornamewithlimits{limsup}}
\def\limitinf{\operatornamewithlimits{liminf}}

\begin{abstract}
Recently, low-complexity and distributed Carrier Sense Multiple
Access (CSMA)-based scheduling algorithms have attracted extensive
interest due to their throughput-optimal characteristics in general
network topologies. However, these algorithms are not well-suited
for serving real-time traffic under time-varying channel conditions
for two reasons: (1) the mixing time of the underlying CSMA Markov
Chain grows with the size of the network, which, for large networks,
generates unacceptable delay for deadline-constrained traffic; (2)
since the dynamic CSMA parameters are influenced by the arrival and
channel state processes, the underlying CSMA Markov Chain may not
converge to a steady-state under strict deadline constraints and
fading channel conditions.

In this paper, we attack the problem of distributed scheduling for
serving real-time traffic over time-varying channels. Specifically,
we consider fully-connected topologies with independently fading
channels (which can model cellular networks) in which flows with
short-term deadline constraints and long-term drop rate requirements
are served. To that end, we first characterize the maximal set of
satisfiable arrival processes for this system and, then, propose a
Fast-CSMA (FCSMA) policy that is shown to be optimal in supporting
any real-time traffic that is within the maximal satisfiable set.
These theoretical results are further validated through simulations
to demonstrate the relative efficiency of the FCSMA policy compared
to some of the existing CSMA-based algorithms.
\end{abstract}

\section{Introduction}
Wireless networks are expected to serve real-time traffic, such as
video or voice applications, generated by a large number of users
over potentially fading channels. These constraints and
requirements, together with the limited shared resources, generate a
strong need for distributed algorithms that can efficiently utilize
the available resources while maintaining high quality-of-service
for the real-time applications. Yet, the strict short-term deadline
constraints and long-term drop rate requirements associated with
most real-time applications complicate the development of provably
good distributed solutions.

In the recent years, there has been an increasing understanding on
the modeling and service of such real-time traffic in wireless
networks (e.g., \cite{kumar09, kumar10, srikant10, eryilmaz10}).
However, existing works in this domain assume centralized
controllers, and hence are not suitable for distributed operation in
large-scale networks. In a separate line of work, it has also been
shown that CSMA-based distributed scheduling (e.g., \cite{jiawal08},
\cite{nisrikant10}, \cite{Ghasri10}, \cite{Rajshah09}) can maximize
long-term average throughput for general wireless topologies.
However, these results also do not apply to strictly
deadline-constrained traffic that we target, since their
throughput-optimality relies: (i) on the convergence time of the
underlying Markov Chain to its steady-state, which grows with the
size of the network; and (ii) on relatively stationary conditions in
which the CSMA parameters do not change significantly over time so
that the instantaneous service rate distribution can stay close to
the stationary distribution. Both of these conditions are violated
in our context: (i) packets of deadline constrained traffic are
likely to be dropped before the CSMA-based algorithm converges to
its steady-state; and (ii) the time-varying fading creates
significant variations on the CSMA parameters, in which case the
instantaneous service rate distribution cannot closely track the
stationary distribution.

While achieving low delay via distributed scheduling in general
topologies is a difficult task (see \cite{shahtse09}), in a related
work \cite{LotMarbach11} that focuses on grid topologies, the
authors have designed an Unlocking CSMA (UCSMA) algorithm with both
maximum throughput and order optimal average delay performance,
which shows promise for distributed scheduling in special
topologies. However, UCSMA also does not directly apply to
deadline-constrained traffic since its measure of delay is on
average. Moveover, it is not clear how existing CSMA or UCSMA
implementations will perform under fading channel conditions.

With this motivation, in this work, we address the problem of
distributed scheduling in fully connected networks (e.g., Cellular
network, WLAN) for serving real-time traffic over independently
fading channels. Our contributions are:

$\bullet$ In Section~\ref{sec:Capacity}, we characterize the maximal
set of satisfiable real-time traffic characteristics as a function
of their drop rate requirements and channel statistics.

$\bullet$ In Section~\ref{sec:FCSMA}, we propose an FCSMA algorithm
that differs from existing CSMA policies in its design principle:
rather than evolving over the set of schedules to reach a favorable
steady-state distribution, the FCSMA policy aims to quickly reach
one of a set of favorable schedules and stick to it for a duration
related to deadline constraints of the application. While the
performance of the former strategy is tied to the mixing-time of a
Markov Chain, the performance of our strategy is tied to the
absorption time, and hence, yields significant advantage for
strictly deadline-constrained flows.

$\bullet$ In Theorem~\ref{thm:opt}, we prove that the FCSMA policy
is optimal in the sense that it can satisfy the deadline and drop
rate requirements for any real-time traffic within the characterized
maximal satisfiable set.

$\bullet$ In Section~\ref{sec:Sim}, we compare the performance of
FCSMA with some of the existing CSMA policies under different
scenarios, both to validate the theoretical claims, and to
demonstrate the performance gains due to our proposed strategy.

\section{System Model}
\label{sec:SystemModel}

We consider a fully-connected wireless network topology where $N$
users contend for data transmission over a single channel that is
independently block fading for each user. We assume that the time
scale of block fading is the same as the duration of the deadline
constraint, and thus uniformly called as a \emph{slot}. We also
assume that all links start transmission at the beginning of each
time slot. We capture the channel fading over link $l$ via $C_l[t],$
which measures the maximum amount of service available in slot $t,$
if scheduled. We assume that $\mathbf{C}[t]=(C_l[t])_{l=1}^N$ are
independently distributed random variables over links and
identically distributed over time. Yet, due to interference
constraints, at most one link can be scheduled for service in each
slot. We use a binary variable $S_l[t]$ to denote whether the link
$l$ is served at slot $t$, where $S_l[t]=1$ if the link $l$ can be
served at slot $t$ and $S_l[t]=0$, otherwise.

Each packet has a delay bound of $1$ time slot, which means that if
a packet cannot be served during the slot it arrives, it will be
dropped. In this context of fully-connected network, we associate
each real-time flow with a link, and hence use these two terms
interchangeably. Let $A_{l}[t]$ denote the number of packets
arriving at link $l$ in slot $t$ that are independently distributed
over links and identically distributed over time with mean
$\lambda_l,$ and $A_l[t]\leq A_{\max}$ for some $A_{\max}<\infty$.
Each link has a maximum allowable drop rate $\rho_{l}\lambda_{l}$,
where $\rho_l\in (0,1)$ is the maximum fraction of packets that can
be dropped at link $l$. For example, $\rho_l=0.1$ means that at most
$10\%$ of packets can be dropped at link $l$ on average. Under above
setup, we define our stochastic control problem (SCP) as follows:
\begin{definition} (SCP)
\begin{eqnarray}
&\displaystyle\maximize_{{\{S[t]\}_{t\geq1}}} & 1 \\
&\text{Subject to   } & \overline{\lambda_l}(1-\rho_l)\leq \underline{\mu_l}, \forall l \label{scp:cons1} \\
&& \sum_{l}S_l[t]\leq 1\label{scp:cons2} \\
&& S_l[t]\in \{0,1\}, \forall l, \forall t\geq 1\label{scp:cons3}
\end{eqnarray}
where
\begin{eqnarray}
\overline{\lambda_l}&=&\limitsup_{T\rightarrow\infty}\frac{1}{T}\sum_{t=1}^{T}\bE[A_l[t]]\\
\underline{\mu_l}&=&\limitinf_{T\rightarrow\infty}\frac{1}{T}\sum_{t=1}^{T}\bE[\min\{S_l[t]C_l[t],A_l[t]\}]
\end{eqnarray}
\end{definition}

In the above maximization problem: (\ref{scp:cons1}) indicates that
the provided average service rates satisfy the drop rate
requirements of the real-time traffic; (\ref{scp:cons2}) indicates
that at most one link is served at each slot.

Normally, it is difficult to solve SCP directly. Instead, we use the
technique in \cite{neely10bk} to introduce a virtual queue $X_l[t]$
for each link $l$ to track the number of dropped packets at slot
$t$. Specifically, the number of packets arriving at virtual queue
$l$ at the end of slot $t$ is denoted as $R_l[t]$, which is equal to
$A_l[t]-\min\{S_l[t]C_l[t],A_l[t]\}$. We use $I_l[t]$ to denote the
service for virtual queue $l$ at the end of the slot $t$ with mean
$\rho_l\lambda_l$, and $I_l[t]\leq I_{\max}$ for some
$I_{\max}<\infty$. Further, we let $U_l[t]$ denote the unused
service for queue $l$ at the end of slot $t$, which is upper-bounded
by $I_{\max}.$
Then, the evolution of
virtual queue is as follows:
\begin{align*}
X_l[t+1]=X_l[t]+R_l[t]-I_l[t]+U_l[t], \quad l=1,\cdots,N.
\end{align*}


In the rest of the paper, we consider the class of stationary
policies $\mathcal{G}$ that select $\mathbf{S}[t]$ as a function of
$(\mathbf{X}[t], \mathbf{A}[t], \mathbf{C}[t]),$ which, then, forms
a Markov Chain. If this Markov Chain is positive recurrent, then the
average drop rate will meet the required constraint automatically
(see \cite{dai95}). 
Accordingly, we call an algorithm \emph{optimal} if it can make this
Markov Chain positive recurrent for any arrival rate vector within
the maximal satisfiable region that we will characterize in the next
section.

\section{FCSMA Algorithm for Throughput Optimality}
In this section, we first study the maximal satisfiable region given
the drop rate and channel statistics. Then, we propose an optimal
FCSMA algorithm.

\subsection{Maximal Satisfiable Region}
\label{sec:Capacity}

Consider the class $\mathcal{G}$ of stationary policies that base
their scheduling decision on the observed vector
$(\mathbf{X}[t],\mathbf{A}[t],\mathbf{C}[t])$ at slot $t$. The next
lemma establishes a condition that is necessary for stabilizing the
system.
\begin{lemma}
If there is a policy $G_{0}\in \mathcal{G}$ that can stabilize the
virtual queue $\mathbf{X}[t]$, then there exist non-negative numbers
$\alpha(\mathbf{a},\mathbf{c};\mathbf{s})$ such that
\begin{align}
\label{lemma1:con1} \sum_{\mathbf{s}\in
\mathcal{S}}\alpha(\mathbf{a},\mathbf{c};\mathbf{s})=1
\end{align}
\begin{scriptsize}
\begin{align}
\label{lemma1:con2}
&\sum_{\mathbf{a}}P_{\mathbf{A}}(\mathbf{a})\sum_{\mathbf{c}}P_{\mathbf{C}}(\mathbf{c})\sum_{\mathbf{s}\in
\mathcal{S}}\alpha(\mathbf{a},\mathbf{c};\mathbf{s})\min\{\mathbf{s}\circ
\mathbf{c},
\mathbf{a}\}>\mathbf{\lambda}\circ(\mathbf{1}-\mathbf{\rho})
\end{align}
\end{scriptsize}
\end{lemma}
where $(\mathbf{A}\circ \mathbf{B})_{i}=A_{i}B_{i}$ denotes Hadamard
product, $P_{\mathbf{A}}(\mathbf{a})=P(\mathbf{A}[t]=\mathbf{a})$
and $P_{\mathbf{C}}(\mathbf{c})=P(\mathbf{C}[t]=\mathbf{c})$.

The proof is almost the same as \cite{tas97} and hence is omitted
here. Note that the left hand side of inequality
($\ref{lemma1:con2}$) is the total average service provided for each
link during one time slot; while $\lambda\circ(1-\rho)$ is the total
average amount of data packets at each link that need to be served.
Thus, to the meet the constraint of drop rate, ($\ref{lemma1:con2}$)
should be satisfied. We define maximal satisfiable region
$\CapReg(\mathbf{\rho})$ as follows:
\begin{align*}
\CapReg(\mathbf{\rho})=\{\mathbf{A}: \exists
\alpha(\mathbf{a},\mathbf{c};\mathbf{s})\geq 0, \text{such that both
} (\ref{lemma1:con1}) \text{ and } (\ref{lemma1:con2})
\text{satisfy}\}
\end{align*}

\subsection{FCSMA algorithm}
\label{sec:FCSMA}

Before we present and analyze our proposed FCSMA algorithm, we
define a set of functions (also see \cite{liery11}) that allows
flexibility in the design and implementation of the algorithm.

$\mathcal{F}:=$ set of non-negative, nondecreasing and
differentiable functions $f(\cdot):\mathbb{R}_{+}\rightarrow
\mathbb{R}_{+}$ with $\displaystyle\lim_{x\rightarrow\infty}
f(x)=\infty$.

$\mathcal{B}:=\{f\in \mathcal{F}$: $\displaystyle
\lim_{x\rightarrow\infty}\frac{f(x+a)}{f(x)}=1$, for any
$a\in\mathbb{R}\}$.

The examples of functions that are in class $\mathcal{B}$ are
$f(x)=\log x$, $f(x)=x$ or $f(x)=e^{\sqrt{x}}$. $f(x)=e^{x}$ is not
in class $\mathcal{B}$.

\begin{definition}[FCSMA Algorithm]
At the beginning of each time slot $t$, each link $l$ independently
generates an exponentially distributed random variable with mean
$f(X_l[t])^{-\min\{C_l[t],A_l[t]\}},$ and starts transmitting after
this random duration unless it senses another transmission before.
The link that grabs the channel transmits its packets until the end
of the slot. If there are no packets awaiting in the link $l$, it
transmits dummy packets to occupy the channel.
\end{definition}
\textit{Remarks}: (1) The absorption time of FCMSA algorithm at slot
$t$ is exponentially distributed with mean
$\frac{1}{\sum_{j=1}^{N}f(X_j[t])^{\min\{C_j[t],A_j[t]\}}}$, which
quickly becomes negligibly small as we demonstrate in next section.

\noindent(2) The parameter of FCSMA policy quickly adapts to arrival
and channel state processes. Due to its fast absorption time, FCSMA
policy yields significant advantages over existing CSMA policies
evolving slowly to the steady-state. In FCSMA, the probability of
serving link $l$ in slot $t$ will be:
\begin{align}
\label{fcsma:def}
\pi_{l}=\frac{f(X_l[t])^{\min\{C_l[t],A_l[t]\}}}{Z}(1-\frac{1}{Z})
\end{align}
where $Z=\sum_{j=1}^{N}f(X_j[t])^{\min\{C_j[t],A_j[t]\}}$. In
equation (\ref{fcsma:def}),
$\frac{f(X_l[t])^{\min\{C_l[t],A_l[t]\}}}{Z}$ is the probability
that link $l$ successfully grabs the channel; while $1-\frac{1}{Z}$
is the average remaining time for serving the packet at slot $t$
given that link $l$ grabs the channel. Let $W^{*}[t]=\max_{l}\log
f(X_{l}[t])\min\{C_{l}[t],A_{l}[t]\}$. The following lemma
establishes the fact that FCSMA policy picks a link with the weight
close to maximum weight with high probability when the maximum
weight $W^{*}[t]$ is large enough.
\begin{lemma}
Given $\epsilon>0$ and $\zeta>0$, $\exists \overline{W}<\infty$,
such that if $W^{*}[t]>\overline{W}$, then FCSMA policy picks a link
$k$ satisfying
\begin{align*}
P\{W_k[t]\geq(1-\epsilon)W^{*}[t]\}\geq1-\zeta
\end{align*}
which also implies
\begin{align}
&\bE[W_k[t]1_{\{W^{*}[t]\geq W\}}|\mathbf{A}[t],\mathbf{C}[t],\mathbf{X}[t]]\nonumber\\
&\geq (1-\epsilon)(1-\zeta)W^{*}[t]1_{\{W^{*}[t]\geq W\}}
\end{align}
\end{lemma}
where $W_{k}[t]=\log f(X_{k}[t])\min\{C_{k}[t],A_{k}[t]\}$.
\begin{IEEEproof}
Define
\begin{align*}
\mathcal{X}&=\{l:\log
f(X_{l}[t])\min\{C_{l}[t],Q_{l}[t]\}<(1-\epsilon)W^{*}[t]\}
\end{align*}
Then,
\begin{align}
\label{lemma2:con} \pi(\mathcal{X})&:=\sum_{l\in
\mathcal{X}}\pi_{l}\nonumber\\
&\leq\sum_{l\in
\mathcal{X}}\frac{\exp(\log f(X_{l}[t])\min\{C_l[t],Q_l[t]\})}{\sum_{j=1}^{n}\exp(\log f(X_{j}[t])\min\{C_{j}[t],Q_{j}[t]\})}\nonumber\\
&<\frac{|\mathcal{X}|\exp((1-\epsilon)W^{*}[t])}{\sum_{j=1}^{n}\exp(\log f(X_{j}[t])\min\{C_{j}[t],Q_{j}[t]\})}\nonumber\\
&\leq\frac{N\exp((1-\epsilon)W^{*}[t])}{\exp(W^{*}[t])}=\frac{N}{\exp(\epsilon
W^{*}[t])}
\end{align}
The first inequality in (\ref{lemma2:con}) follows the fact that
$1-\frac{1}{Z}\leq1$. Thus, $\exists\overline{W}<\infty$ such that
$W^{*}[t]>\overline{W}$ implies $\pi(\mathcal{X})<\zeta$.
\end{IEEEproof}

Under certain conditions for the function $f$, we can establish the
optimality of FCSMA algorithm.
\begin{theorem} \label{thm:opt}
FCSMA is optimal if $\log f\in \mathcal{B}$ and $f(0)\geq1$.
\end{theorem}
\begin{IEEEproof}
See the Appendix for the proof.
\end{IEEEproof}
\textit{Remarks}: The optimality of FCSMA is preserved even when the
slope of function $f$ is low, which is easier to be implemented in
practice.
\section{Simulation Results}
\label{sec:Sim}

In this section, we perform simulations to validate the optimality
of the proposed FCSMA policy with deadline constraint $1$ time slot
in both fading and non-fading channels. In the simulation, there are
$N=10$ links. All links require that the maximum fraction of
dropping packets cannot exceed $\rho=0.2$. The number of arrivals in
each slot follows Bernoulli distribution. For the simulations of a
fading channel, all links suffer from the ON-OFF channel fading
independently with probability $p=0.9$ that the channel is available
in each time slot. Under this setup, we can use the same technique
in paper \cite{taseph93} to get the maximal satisfiable region:
$\Gamma=\{\lambda:N(1-\rho)\lambda<1-(1-p\lambda)^N\}$. Through
numerical calculation, we can get $\lambda<0.051$ in non-fading
channel and $\lambda<0.03$ in fading channel. We compare our
proposed FCSMA policy with $f(x)=e^x$ with QCSMA algorithm
\cite{nisrikant10} with the weight $X_l[t]\min\{C_l[t], A_l[t]\}$
(In our setup, QCSMA algorithm with the weight
$\log\log(X_l[t]\min\{C_l[t], A_l[t]\}+e)$ has much worse
performance than that with $X_l[t]\min\{C_l[t], A_l[t]\}$). To that
end, we divide each time slot into $M$ mini-slots. In FCSMA policy,
if the link contends for the channel successfully, it will occupy
that channel in the rest of time slot; while in QCSMA policy, each
link contends for the channel and transmits the data in $1$
mini-slot. Here, we don't consider the overhead that the QCSMA
policy needs to contend for the channel, which will greatly degrade
its performance.

\begin{figure}[h]
\center
\includegraphics[scale=0.31]{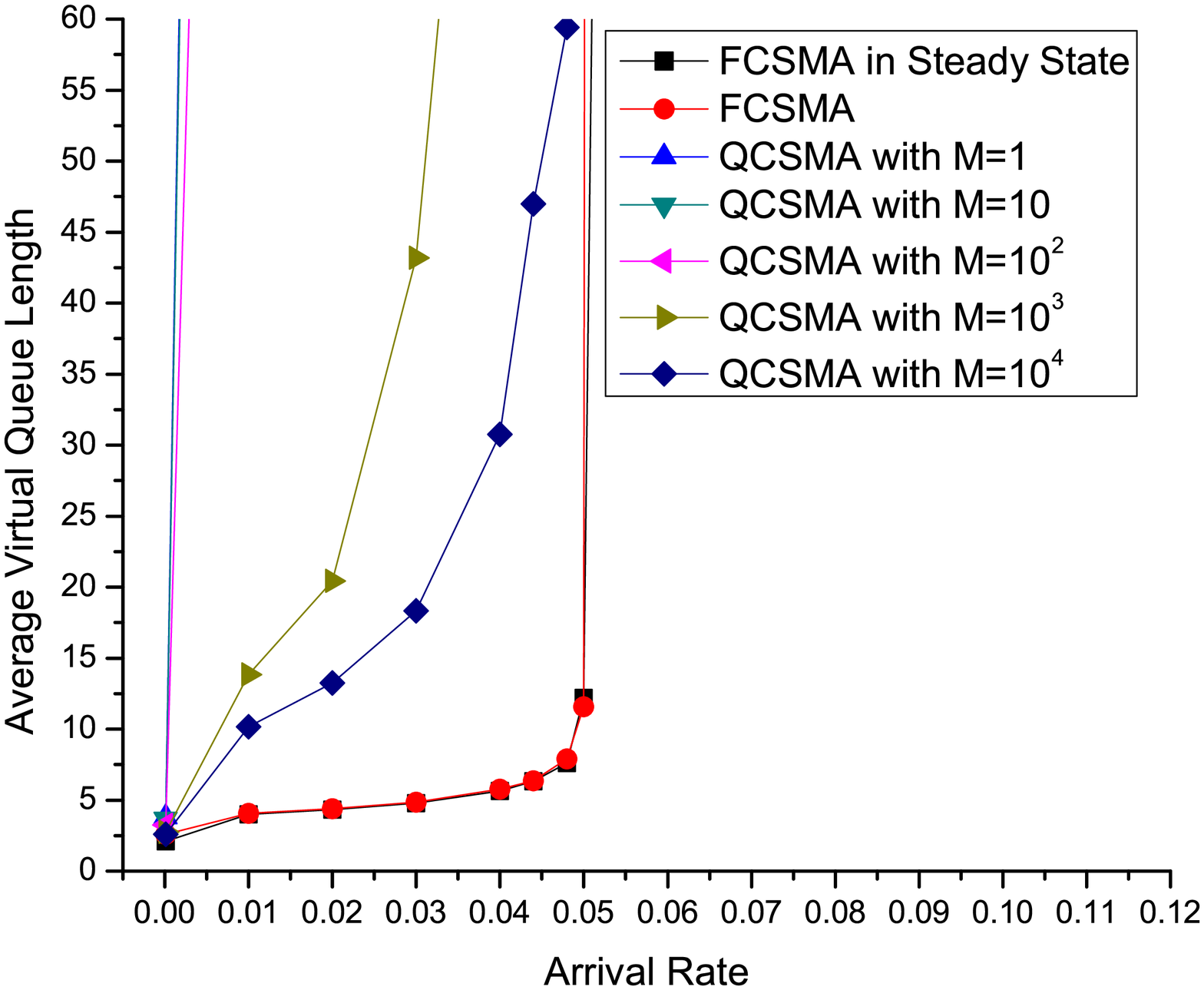}
\caption{Performance of FCSMA and QCSMA over non-fading
channel}\label{sim:fig1}
\end{figure}

\begin{figure}[h]
\center
\includegraphics[scale=0.31]{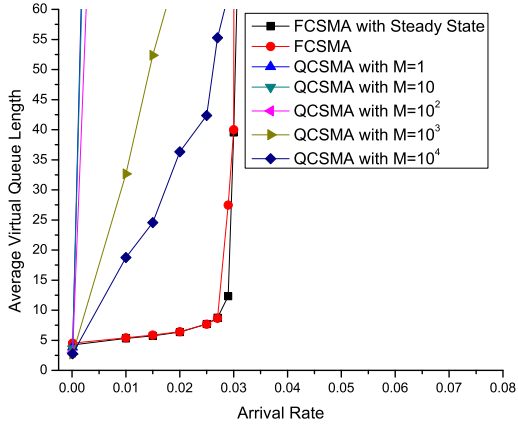}
\caption{Performance of FCSMA and QCSMA over fading
channel}\label{sim:fig2}
\end{figure}

From Figure \ref{sim:fig1} and \ref{sim:fig2}, we can observe that
the average virtual queue length grows very fast under the QCSMA
policy with $M=1$ while the average queue length of FCSMA always
stays at a low level. The reason for the poor performance of QCSMA
scheme in deadline-constrained application is that the underlying
Markov chain is controlled by the arrival and channel state
processes. If the running time of QCSMA policy has the same time
scale with the deadline of the packet, this Markov chain cannot
converge to the steady-state. However, FCMSA policy can quickly lock
into one state and exhibits good performance, which is shown in
Theorem $1$ to be optimal if we carefully choose the parameters. In
addition, as $M$ increases, the performance of QCSMA improves. The
reason is that the underlying Markov chain has enough time to
converge to the steady-state and thus yields better performance.
Recall that FCSMA policy waits for random duration before accessing
the channel, this random duration can be arbitrarily small when the
number of links increases and the virtual queue length is high. We
can see from simulations that FCSMA policy has almost the same
performance as that in steady state.


\section{Conclusions}
In this paper, we first characterized the maximal satisfiable set of
arrival processes given the drop rate and channel statistics and
then proposed a provably optimal distributed FCSMA policy for
scheduling deadline-constrained traffic over fading channel. We
validated the performance of FCSMA policy by comparing it with
existing CSMA policies through simulations. We assumed that the time
scale of channel fading is the same as the duration of the deadline
constraint, which is not always the case in practical wireless
networks. We will relax this assumption in our future work. Also, we
will try to explore scheduling algorithms for real-time traffic over
fading channel in multi-hop network topologies.

\section{Acknowledgement}
This work was supported in part by DTRA Grant HDTRA 1-08-1-0016, and
NSF Awards: CAREER-CNS-0953515 and CCF-0916664.

\section{Appendix \\Proof of Theorem  $1$}
\begin{IEEEproof}
Let $g(x)=\log f(x)$. Consider the Lyapunov function
$V(\mathbf{X}):=\sum_{l=1}^{N}h(X_{l})$, where $h'(x)=g(x)$. Then
\begin{align*}
\Delta V:&=\bE\left[V(\mathbf{X}[t+1])-V(\mathbf{X}[t])
        |\mathbf{X}[t]=\mathbf{X}\right]\\
         &=\sum_{l=1}^{N}\bE\left[(h(X_{l}[t+1])-h(X_{l}[t]))|\mathbf{X}[t]=\mathbf{X}\right]
\end{align*}
By the mean-value theorem, we have $
h(X_{l}[t+1])-h(X_{l}[t])=g(X_{l}')(X_{l}[t+1]-X_{l}[t])=g(X_{l}')(R_{l}[t]-I_{l}[t]+U_{l}[t])$,
where $X_{l}'$ lies between $X_{l}[t]$ and $X_{l}[t+1]$. Hence, we
get
\begin{align*}
\Delta
V=&\sum_{l=1}^{N}\bE\left[g(X_{l}')(R_{l}[t]-I_{l}[t]+U_{l}[t])|\mathbf{X}[t]=\mathbf{X}\right]
\nonumber \\
=&\underbrace{\sum_{l=1}^{N}\bE\left[g(X_{l}')U_{l}[t]|\mathbf{X}[t]\right]}_{=:\Delta
V_1}+\underbrace{\sum_{l=1}^{N}\bE\left[g(X_{l}')(R_{l}[t]-I_{l}[t])|\mathbf{X}[t]\right]}_{=:\Delta
V_2}
\end{align*}
For $\Delta V_1$, if $X_{l}[t]=X_{l}\geq I_{\max}$, then
$U_{l}[t]=0$. If $X_{l}[t]=X_{l}<I_{\max}$, then $U_{l}[t]\leq
I_{\max}$. But in this case, $X_{l}[t+1]\leq (I_{\max}+A_{\max})$.
Hence, $g(X_{l}')\leq g(I_{\max}+A_{\max})<\infty$. Thus,
\begin{align}
\Delta
V_1&=\sum_{l=1}^{N}\bE\left[g(R_{l})U_{l}[t]\mathbf{1}_{\{X_{l}<I_{\max}\}}|\mathbf{X}[t]=\mathbf{X}\right]\nonumber\\
&\leq NI_{\max}g(I_{\max}+A_{\max})
\end{align}
where $\mathbf{1}_{\{\cdot\}}$ is the indicator function.

Next, let's focus on $\Delta V_2$. We know that
$g(X_{l}')=g(X_{l}[t]+a_{l})$ ($|a_{l}|\leq A_{\max}$). According to
the definition of function $g\in \mathcal{B}$, given $\beta>0$,
there exists $M>0$, such that for any $X_{l}[t]=X_{l}>M$, we have
$\left|\frac{g(X_{l}')}{g(X_{l})}-1\right|<\beta$, that is,
\begin{align}
(1-\beta)g(X_{l})<g(X_{l}')<(1+\beta)g(X_{l})
\end{align}
Thus, we have
\begin{align}
\label{th1:ineq}
&g(X_{l}')(R_{l}[t]-I_{l}[t])\nonumber\\
=&g(X_{l}')\left[(R_{l}[t]-I_{l}[t])_{+}-(R_{l}[t]-I_{l}[t])_{-}\right]\nonumber \\
<&(1+\beta)g(X_{l})(R_{l}[t]-I_{l}[t])_{+}\nonumber\\
&-(1-\beta)g(X_{l})(R_{l}[t]-I_{l}[t])_{-}
\nonumber \\
=&g(X_l)(R_l[t]-I_l[t])+\beta
g(X_{l})\left|R_{l}[t]-I_{l}[t]\right|\nonumber\\
\leq& g(X_l)(R_l[t]-I_l[t])+\beta A_{\max}g(X_{l})
\end{align}
where $(x)_{+}=\max\{x,0\}$, $(x)_{-}=-\min\{x,0\}$ and
$|R_{l}[t]-I_{l}[t]|\leq |A_{l}[t]|\leq A_{\max}$. Thus, we divide
$\Delta V_2$ into two parts:
\begin{align*}
&\Delta V_2\nonumber=\underbrace{\sum_{l=1}^{N}\bE\left[g(X_{l}')(R_{l}[t]-I_{l}[t])\mathbf{1}_{\{X_{l}>M\}}|\mathbf{X}[k]=\mathbf{X}\right]}_{=:\Delta V_3}\nonumber\\
&+\underbrace{\sum_{l=1}^{N}\bE\left[g(X_{l}')(R_{l}[t]-I_{l}[t])\mathbf{1}_{\{X_{l}\leq
M\}}|\mathbf{X}[t]=\mathbf{X}\right]}_{=:\Delta V_4}
\end{align*}
For $\Delta V_3$, by using (\ref{th1:ineq}), we have
\begin{align*}
\Delta V_3\leq&\sum_{l=1}^{N}\bE\left[g(X_{l})(R_{l}[t]-I_{l}[t])\mathbf{1}_{\{X_{l}>M\}}|\mathbf{X}[t]=\mathbf{X}\right]\nonumber\\
&+\sum_{l=1}^{N}\beta A_{\max}g(X_{l})\mathbf{1}_{\{X_{l}>M\}}\nonumber\\
=&\underbrace{\sum_{l=1}^{N}\bE[g(X_{l})(A_{l}[t]-I_{l}[t])\mathbf{1}_{\{X_{l}>M\}}|\mathbf{X}[t]=\mathbf{X}]}_{=:L_1}\nonumber\\
-&\underbrace{\bE[\sum_{l=1}^{N}W^{F}_{l}[t]1_{\{X_l>M\}}|\mathbf{X}[t]]}_{=:L_2}+\sum_{l=1}^{N}\beta
A_{\max}g(X_{l})\mathbf{1}_{\{X_{l}>M\}}
\end{align*}
where $W_{l}^{F}[t]=\log
f(X_{l}[t])\min\{C_{l}[t]S_l^{F}[t],A_{l}[t]\}$ and
$\mathbf{S}^{F}[t]$ denotes the schedule chosen by FCSMA with
$S^{F}_{k}[t]=1$. Next, we will explore the upper bound of $L_1$ by
using Lemma $1$ and give the lower bound of $L_2$ by the Lemma $2$.

First, let's focus on $L_1$. By Lemma $1$, there exist non-negative
numbers $\alpha(a,c;s)$ satisfying (\ref{lemma1:con1}) and for a
$\delta>0$ small enough, we have
\begin{align}
\label{th1:con1}
&\sum_{\mathbf{a}}P_\mathbf{A}(\mathbf{a})\sum_{\mathbf{c}}P_\mathbf{C}(\mathbf{c})\sum_{\mathbf{s}\in
\mathcal{S}}\alpha(\mathbf{a},\mathbf{c};\mathbf{s})\min\{s_lc_l, a_l\}\nonumber\\
&\geq \lambda_l(1-\rho_l)+\delta
\end{align}
Let $W_l=g(X_l)\min\{s_lc_l,a_l\}$. In the following proof, we can
also write the maximum weight $W^{*}[t]=\sum_{l=1}^{N}W^{*}_{l}[t]$,
where $W^{*}_l[t]=\log
f(X_{l}[t])\min\{C_{l}[t]S_l^{*}[t],A_{l}[t]\}$ and optimal schedule
$\mathbf{S}^{*}[t]=\arg\max_{\mathbf{S}\in
\mathcal{S}}\sum_{l=1}^{N}W^{*}_{l}[t]$. By using (\ref{th1:con1}),
we have
\begin{align}
\label{th1:L1}
&L_1=\sum_{l=1}^{N}g(X_l)\lambda_l(1-\rho_l)1_{\{X_l>M\}}\nonumber\\
&\leq\sum_{\mathbf{a}}P_\mathbf{A}(\mathbf{a})\sum_{\mathbf{c}}P_\mathbf{C}(\mathbf{c})\sum_{\mathbf{s}\in
\mathcal{S}}\alpha(\mathbf{a},\mathbf{c};\mathbf{s})\sum_{l=1}^{N}W_{l}1_{\{X_l>M\}}\nonumber\\
&-\delta\sum_{l=1}^{N}g(X_l)1_{\{X_l>M\}}\nonumber\\
=&\sum_{\mathbf{a}}P_\mathbf{A}(\mathbf{a})\sum_{\mathbf{c}}P_\mathbf{C}(\mathbf{c})\sum_{\mathbf{s}\in
\mathcal{S}}\alpha(\mathbf{a},\mathbf{c};\mathbf{s})\sum_{l=1}^{N}W_{l}1_{\{X_l>M,W^{*}[t]>\overline{W}\}}\nonumber\\
+&\sum_{\mathbf{a}}P_\mathbf{A}(\mathbf{a})\sum_{\mathbf{c}}P_\mathbf{C}(\mathbf{c})\sum_{\mathbf{s}\in
\mathcal{S}}\alpha(\mathbf{a},\mathbf{c};\mathbf{s})\sum_{l=1}^{N}W_{l}1_{\{X_l>M,W^{*}[t]\leq \overline{W}\}}\nonumber\\
-&\delta\sum_{l=1}^{N}g(X_l)1_{\{X_l>M\}}\nonumber\\
\leq&\sum_{\mathbf{a}}P_\mathbf{A}(\mathbf{a})\sum_{\mathbf{c}}P_\mathbf{C}(\mathbf{c})\sum_{\mathbf{s}\in
\mathcal{S}}\alpha(\mathbf{a},\mathbf{c};\mathbf{s})\sum_{l=1}^{N}W_{l}1_{\{X_l>M,W^{*}[t]>\overline{W}\}}\nonumber\\
&+\overline{W}-\delta\sum_{l=1}^{N}g(X_l)1_{\{X_l>M\}}
\end{align}
Second, let's consider $L_2$. Since
\begin{align*}
&(1-\epsilon)(1-\zeta)\bE[\sum_{l=1}^{N}W^{*}_{l}[t]1_{\{X_l>M, W^{*}[t]>\overline{W}\}}|\mathbf{X}[t]=\mathbf{X}]\nonumber\\
&\leq(1-\epsilon)(1-\zeta)\bE[\sum_{l=1}^{N}W^{*}_{l}[t]1_{\{W^{*}[t]>\overline{W}\}}|\mathbf{X}[t]=\mathbf{X}]\nonumber\\
&\leq\bE[\sum_{l=1}^{N}W^{F}_{l}[t]1_{\{W^{*}[t]>\overline{W}\}}|\mathbf{X}[t]=\mathbf{X}] (\text{By Lemma 2})\nonumber\\
&=\bE[\sum_{l=1}^{N}W^{F}_{l}[t]1_{\{X_l>M,W^{*}[t]>\overline{W}\}}|\mathbf{X}[t]=\mathbf{X}]\nonumber\\
&+\bE[\sum_{l=1}^{N}W^{F}_{l}[t]1_{\{X_l\leq M,W^{*}[t]>\overline{W}\}}|\mathbf{X}[t]=\mathbf{X}]\nonumber\\
&\leq\bE[\sum_{l=1}^{N}W^{F}_{l}[t]1_{\{X_l>M,W^{*}[t]>\overline{W}\}}|\mathbf{X}[t]=\mathbf{X}]+NA_{\max}g(M)
\end{align*}
$L_2$ becomes
\begin{align}
\label{th1:L2}
L_2&\geq\bE[\sum_{l=1}^{N}W^{F}_{l}[t]1_{\{X_l>M,W^{*}[t]>\overline{W}\}}|\mathbf{X}[t]]\nonumber\\
&\geq(1-\epsilon)(1-\zeta)\bE[\sum_{l=1}^{N}W^{*}_{l}[t]1_{\{X_l>M,
W^{*}[t]>\overline{W}\}}|\mathbf{X}[t]]\nonumber\\
&-NA_{\max}g(M)
\end{align}
Thus, by using (\ref{th1:L1}) and (\ref{th1:L2}), $\Delta V_3$
becomes
\begin{align}
&\Delta V_3\leq
\sum_{\mathbf{a}}P_A(\mathbf{a})\sum_{\mathbf{c}}P_C(\mathbf{c})\sum_{\mathbf{s}\in
\mathcal{S}}\alpha(\mathbf{a},\mathbf{c};\mathbf{s})\sum_{l=1}^{N}W_{l}1_{\{X_l>M,W^{*}[t]>\overline{W}\}}\nonumber\\
&-\bE[\sum_{l=1}^{N}W^{*}_{l}[t]1_{\{X_l>M,
W^{*}[t]>\overline{W}\}}|\mathbf{X}[t]]\nonumber\\
&+(\epsilon+\zeta-\epsilon\zeta)\bE[\sum_{l=1}^{N}W^{*}_{l}[t]1_{\{X_l>M, W^{*}[t]>\overline{W}\}}|\mathbf{X}[t]]\nonumber\\
&+\overline{W}+NA_{\max}g(M)-\delta\sum_{l=1}^{N}g(X_l)1_{\{X_l>M\}}\nonumber\\
&+\beta A_{\max}\sum_{l=1}^{N}g(X_l)1_{\{X_l>M\}}
\end{align}
Since
\begin{align}
\label{th1:con2}
&\sum_{\mathbf{a}}P_A(\mathbf{a})\sum_{\mathbf{c}}P_C(\mathbf{c})\sum_{\mathbf{s}\in
\mathcal{S}}\alpha(\mathbf{a},\mathbf{c};\mathbf{s})\sum_{l=1}^{N}W_l-\bE[\sum_{l=1}^{N}W^{*}_{l}[t]|\mathbf{X}[t]]\nonumber\\
=&\sum_{\mathbf{a}}P_A(\mathbf{a})\sum_{\mathbf{c}}P_C(\mathbf{c})\sum_{\mathbf{s}\in
\mathcal{S}}\alpha(\mathbf{a},\mathbf{c};\mathbf{s})\sum_{l=1}^{N}W_l\nonumber\\
&-\sum_{\mathbf{a}}P_A(\mathbf{a})\sum_{\mathbf{c}}P_C(\mathbf{c})\sum_{\mathbf{s}\in
\mathcal{S}}\alpha(\mathbf{a},\mathbf{c};\mathbf{s})\sum_{l=1}^{N}W^{*}_l\nonumber\\
&\leq0
\end{align}
Thus, we have
\begin{align}
\label{th1:delv3_1}
&\sum_{\mathbf{a}}P_A(\mathbf{a})\sum_{\mathbf{c}}P_C(\mathbf{c})\sum_{\mathbf{s}\in
\mathcal{S}}\alpha(\mathbf{a},\mathbf{c};\mathbf{s})\sum_{l=1}^{N}W_{l}1_{\{X_l>M,W^{*}[t]>\overline{W}\}}\nonumber\\
&\leq\sum_{\mathbf{a}}P_A(\mathbf{a})\sum_{\mathbf{c}}P_C(\mathbf{c})\sum_{\mathbf{s}\in
\mathcal{S}}\alpha(\mathbf{a},\mathbf{c};\mathbf{s})\sum_{l=1}^{N}W_{l}\nonumber\\
&\leq\bE[\sum_{l=1}^{N}W^{*}_{l}[t]|\mathbf{X}[t]=\mathbf{X}](\text{By using (\ref{th1:con2})})\nonumber\\
&=\bE[\sum_{l=1}^{N}W^{*}_{l}[t]1_{\{X_l>M\}}|\mathbf{X}[t]]+\bE[\sum_{l=1}^{N}W^{*}_{l}[t]1_{\{X_l\leq
M\}}|\mathbf{X}[t]]\nonumber\\
&\leq\bE[\sum_{l=1}^{N}W^{*}_{l}[t]1_{\{X_l>M,W^{*}[t]>\overline{W}\}}|\mathbf{X}[t]=\mathbf{X}]\nonumber\\
&+\bE[\sum_{l=1}^{N}W^{*}_{l}[t]1_{\{X_l>M,W^{*}[t]\leq \overline{W}\}}|\mathbf{X}[t]=\mathbf{X}]+NA_{\max}g(M)\nonumber\\
&\leq\bE[\sum_{l=1}^{N}W^{*}_{l}[t]1_{\{X_l>M,W^{*}[t]>\overline{W}\}}|\mathbf{X}[t]]+\overline{W}+NA_{\max}g(M)
\end{align}
In addition, we have
\begin{align}
\label{th1:delv3_2} &\bE[\sum_{l=1}^{N}W^{*}_{l}[t]1_{\{X_l>M,
W^{*}[t]>\overline{W}\}}|X[t]=X]\nonumber\\
&\leq\bE[\sum_{l=1}^{N}W^{*}_{l}[t]|X[t]=X]\nonumber\\
&\leq A_{\max}\sum_{l=1}^{N}g(X_l)\nonumber\\
&=A_{\max}\sum_{l=1}^{N}g(X_l)1_{\{X_l>M\}}+A_{\max}\sum_{l=1}^{N}g(X_l)1_{\{X_l\leq
M\}}\nonumber\\
&\leq A_{\max}\sum_{l=1}^{N}g(X_l)1_{\{X_l>M\}}+NA_{\max}g(M)
\end{align}
then, by using (\ref{th1:delv3_1}) and (\ref{th1:delv3_2}), we have
\begin{align}
\Delta V_3&\leq -\gamma\sum_{l=1}^{N}g(X_l)1_{\{X_l>M\}}+D_1
\end{align}
where
$D_1=2\overline{W}+(2+\epsilon+\zeta-\epsilon\zeta)NA_{\max}g(M)$
and $\gamma=\delta-\beta
A_{\max}-A_{\max}(\epsilon+\zeta-\epsilon\zeta)$. We can choose
$\beta, \epsilon, \zeta$ small enough such that $\gamma>0$.

For $\Delta V_4$, we have
\begin{align*}
\Delta V_4
&\leq\sum_{l=1}^{N}\bE\left[g(X_{l}')R_{l}[t]|\mathbf{X}[t]=\mathbf{X}\right]\mathbf{1}_{\{X_{l}\leq
M\}}\nonumber\\
&\leq\sum_{l=1}^{N}\bE\left[g(X_{l}')A_{l}[t]|\mathbf{X}[t]=\mathbf{X}\right]\mathbf{1}_{\{X_{l}\leq
M\}}\nonumber\\
&\leq NA_{\max}g(M+A_{\max})
\end{align*} Thus, we get
\begin{align}
\label{th1:result} \Delta
V&<-\gamma\sum_{l=1}^{N}g(X_{l})\mathbf{1}_{\{X_{l}>M\}}+D\nonumber\\
&\leq-\gamma\sum_{l=1}^{N}g(X_{l})+E
\end{align}
where
$D:=NI_{\max}g(I_{\max}+A_{\max})+D_1+NA_{\max}g(M+A_{\max})<\infty$
and $E:=D+N\gamma g(M)$. Hence, by the Lyapunov Drift theorem
\cite{neely10bk}, we have
$\limsup_{T\rightarrow\infty}\frac{1}{T}\sum_{t=0}^{T-1}\sum_{l=1}^{N}\bE[g(X_l[t])]\leq
\frac{E}{\gamma}<\infty$, which implies stability-in-the mean and
thus the Markov Chain is positive recurrent \cite{meyn92}.
\end{IEEEproof}

\begin{spacing}{}
\bibliographystyle{abbrv}
\bibliographystyle{IEEEtran}
\bibliography{refs}
\end{spacing}

\end{document}